\documentstyle[11pt,psfig,aaspp4]{article}

\begin{document}

\title{The structure of dark matter halos in hierarchical clustering
theories. }

\author{Kandaswamy Subramanian}

\affil{ National Centre for Radio Astrophysics,
Tata Institute of Fundamental Research, Poona University Campus,
Ganeshkhind, Pune 411 007, India.}
\bigskip
\author{ Renyue Cen and Jeremiah P. Ostriker }

\affil{Princeton University Observatory, Peyton Hall,
Princeton, New Jersey 08544, USA.}

\begin{abstract}

During hierarchical clustering, smaller masses generally
collapse earlier than larger masses and so are denser on 
the average. The core of a small mass halo could be dense 
enough to resist disruption and survive undigested, when it 
is incorporated into a bigger object. We explore the 
possibility that a nested sequence of undigested cores in 
the center of the halo, which have survived the hierarchical, 
inhomogeneous collapse to form larger and larger objects, 
determines the halo structure in the inner regions.
For a flat universe with $P(k) \propto k^n$,
scaling arguments then suggest that the core density profile is,
$\rho \propto r^{-\alpha}$ with $\alpha = (9+3n)/(5+n)$.
For any $n < 1$, the signature of undigested cores is a 
core density profile shallower than
$\rho \propto 1/r^2$ and dependent on the power spectrum.
For typical objects formed from a CDM like power spectrum
the effective value of $n$ is close to -2 and thus 
$\alpha$ could typically be near 1, the NFW (see text) value.
Also velocity dispersions should deviate from a constant value to 
decrease with decreasing radius in the core.
But whether such behaviour obtains depends on detailed dynamics.

We first examine the dynamics using a fluid approach to the 
self-similar collapse solutions for the dark matter phase 
space density, including the effect of velocity dispersions. 
We highlight the importance of tangential velocity dispersions 
to obtain density profiles shallower than $1/r^2$ in the core 
regions. If tangential velocity dispersions in the core 
are constrained to be less than the radial dispersion,
a cuspy core density profile shallower than $1/r$ cannot obtain, 
in self-similar collapse. 
We then briefly look at the profiles of the outer halos in
low density cosmological models where the total halo
mass is convergent. We find a limiting $r^{-4}$  outer profile
for the open case and a limiting outer profile for 
the $\Lambda$ dominated case, which at late times has
the form $[1 - (r/\bar r_{\lambda})^{-3\epsilon}]^{1/2}$, 
where $3\epsilon$ is the logarithmic slope of the 
initial density profile.
Finally, we analyze a suite of dark halo density and 
velocity dispersion profiles obtained in cosmological N-body
simulations of models with $n= 0, -1$ and $-2$. We find that
the core-density profiles of dark halos, show considerable scatter
in their properties, but nevertheless do appear to
reflect a memory of the initial power spectrum, with 
steeper initial spectra producing flatter core profiles.
These results apply as well for low density cosmological
models ($\Omega_{matter} = 0.2 - 0.3$), since high density
cores were formed early where $\Omega_{matter} \approx 1$.

\end{abstract}

\section{Introduction}

In conventional pictures for the growth of structure,
galaxies and clusters are thought to originate from the growth 
of small density fluctuations due to gravitational
instability, in a universe dominated by dark matter. 
In hierarchical clustering models, like the cold dark matter (CDM) models,
small mass clumps of dark matter form first and gather into larger 
and larger masses subsequently. The structure of these dark matter clumps,
which we will refer to as "halos", is likely to be 
related to how the halos formed, the initial spectrum
of the density fluctuations and to the underlying cosmology.

Much of the early work on the structure of dark halos 
concentrated on their density profiles in the outer regions,
especially in the context of understanding the flat rotation
curves of disk galaxies. The secondary infall paradigm 
introduced by Gunn and Gott (1972) and subsequently amplified
by Fillmore and Goldreich (FG, 1984) and Bertschinger (B85, 1985)
suggests that gravitational collapse around a seed 
perturbation will generically lead to divergent extended halos which
produce a nearly flat rotation curve in the outer regions for
the case $\Omega_{matter} \equiv \Omega_m =1$. 
The Gunn-Gott picture would lead
to steep convergent profiles in the outer regions for low
density ($\Omega_{m} < 1$) universes. If current estimates
for the global density parameter are correct ($\Omega_m = 0.3 \pm 0.1$),
then the high density core ($\rho_{core}/<\rho> > 10^3$ ) were formed
early enough so that the $\Omega_m = 1$ picture effectively
applies. But the outer halos represent the current, low density 
universe.

The nature of the density profiles of dark halos in the inner regions
is also of importance from several points of view.
The structure of dark halo cores determines the efficiency
of gravitational lensing by the galactic and cluster halos, 
the X- ray emissivity of clusters and
galactic rotation curves in the inner regions.
These properties of galactic and cluster halos can be well 
constrained by observations.  So, if the core density profile of dark
halos are fossils which do depend on some of the properties of structure 
formation models, like their initial power spectrum, 
one would have a useful observational handle on these properties.
It is therefore necessary to understand what determines 
the nature of the density profiles of dark matter halos, 
and their cores, {\it ab initio}. We discuss this issue here.

Further, Navarro, Frenk and White (NFW) (1995, 96, 97)
have proposed from their N-body simulations, 
that dark matter halos in hierarchical 
clustering scenarios develop a universal density profile,
regardless of the scenario for structure formation or cosmology.
The NFW profile has an inner cuspy form with the density
$\rho \propto r^{-1}$ and an outer envelope of
the form $\rho \propto r^{-3}$. There does not
appear to be any reason, apriori, why halo density profiles should
prefer such a form, 
but empirically, several investigators have found that the NFW
profile provides a moderately good fit to numerical simulations
( Cole and Lacey 1996,
Tormen, Bouchet $\&$ White 1997, Huss, Jain $\&$ Steinmetz 1997, 1999,
Thomas {\it et al.}, 1998). Recently, though, high resolution 
simulations of cluster formation in a CDM model, 
by Moore {\it et al.} (1998), yielded a core 
density profile $\rho(r) \propto r^{-1.4}$, 
shallower than $r^{-2}$, but steeper than the $r^{-1}$ form 
preferred by NFW, consistent with the earlier high resolution
work of Xu (1995). A similar result was also found earlier
by Fukushige and Makino (1997). (For small mass halo 
cores on the other hand, Kravtsov {\it et al.} (1998), find
an average core density profile, shallower than the NFW form ).
Xu (1995) also found that there was a large scatter in the 
the logarithmic slope of halo density profiles in both the
core and outer regions. One motivation of our work 
was to examine this issue on general theoretical grounds; 
while at the same time check in some of our own
numerical experiments, the properties of dark halo
density and velocity dispersion profiles.

In the next section we discuss the processes which may determine the
halo density profile and consider the role
of undigested cores in setting the structure of dark halos cores.
For a flat universe with $P(k) \propto k^n$,
scaling arguments suggest that 
$\rho \propto r^{-\alpha}$ with $\alpha = \alpha_n = (9+3n)/(5+n)$.
As an aside we note here that for popular cosmological
models $n \approx -2$, in the appropriate range of 
wavelengths, giving $\alpha = 1$, the NFW value.
But whether such a scaling law indeed obtains depends on the 
detailed dynamics.

In order to explore the dynamical issues, we consider first
self similar collapse of a single spherically symmetric density
perturbation, in a scale free universe. We introduce a fluid
approach for analyzing this problem, in Section 3. 
We highlight the importance of tangential velocity dispersions to 
obtain density laws shallower than $1/r^2$ in the core regions.
In a companion paper (Subramanian, 1999, S99 henceforth), one of us (KS) 
considers these self-similar collapse solutions in greater detail,
by deriving and solving numerically the scaled moment equations
for such a collapse, including the effect of
tangential velocity dispersions. 
 
In Section 4 we analyze, following the Gunn-Gott paradigm,
the outer profiles expected in low density universes, where
an outer profile steeper than $r^{-3}$ must obtain.
In Section 5, we analyze dark halo density and velocity dispersion
profiles obtained in cosmological N-body
simulations of models with $n= 0, -1$ and $-2$. We show that
the core-density profiles of dark halos, show some scatter
in their properties, but nevertheless do appear to
reflect a memory of the initial power spectrum. 
The final Section discusses the results and presents our conclusions.

\section{The density profiles of dark halos}

To fix ideas, let us limit ourself initially, to the Einstein de-Sitter
universe, with $\Omega =1$. 
This is almost certainly not the correct cosmological model,
but it provides a convenient context within which to  discuss
the formation of structure and it is likely
to be a very good approximation at epochs $(1+z)>\Omega_m^{-1}$
at which the cores of familiar objects have formed.
Let us also assume that the
Fourier space power spectrum of density fluctuations is a power law,
$P(k) = A k^n$, where the spectral index n lies between the limits
$-3  <  n < 1$. In this case structure grows hierarchically
with small scales going non-linear first and larger and larger
mass scales going non-linear at progressively later times.
For such a scale free universe, 
all properties of dark matter distributions at 
each epoch are just a 
scaled version of those in previous epochs i.e. the universe is
self-similar. What would decide the density profile 
of a dark matter halo in such a cosmological setting?

It is likely that at least three processes are important. 
Firstly, when some mass scale decouples from the general
Hubble expansion and collapses in an inhomogeneous
fashion to form a dark matter halo, 
the changing gravitational potential and
phase mixing will cause some amount of violent relaxation
or " virialisation" to occur.
A general constraint on the  
equilibrium distribution of such a halo will be set by
energy and mass conservation together with scaling laws which obtain in
a hierarchical clustering scenario. 

Secondly, in the cosmological context, a collapsed mass is not
isolated and will therefore continue to accrete surrounding
material, as long as matter dominates the energy density.
Such a secondary infall onto the collapsed 
halo will alter/determine its structure in the outer
regions. 

We emphasize here a third process: 
When any mass scale collapses, in a hierarchical 
theory, it will already contain a dominant smaller mass dark halo
which collapsed earlier, and is therefore denser
on the average. Typically the 
core of such a smaller mass halo, is dense enough to resist
disruption and survive undigested, when it is
incorporated into the bigger object. 
A nested sequence of 
undigested cores in the center of the halo,
which have survived the hierarchical 
inhomogeneous collapse to form larger and larger objects, 
could thus determine the halo structure in the inner regions.
We illustrate this idea schematically in Figure 1.

Suppose a halo of mass M collapses to form a "virialised" 
object with a characteristic density $\rho_0$ and core radius $r_c$.
For $P(k) \propto k^n$, simple standard scaling arguments 
using linear theory (cf. Peebles 1980, Padmanabhan and Subramanian 1992,
Padmanabhan 1993), predict that $\rho_0$ and 
$r_c$, scale with mass $M$ as 
\begin{equation}
\rho_0(M) \propto  M^{-(n+3)/2}  ;\qquad
r_c(M) \propto  M^{(n+5)/6} ;\qquad
\rho_0 \propto r_c^{-(9+3n)/(5+n)}
\label{scale}
\end{equation}
So, in the above sequential collapse to form larger and larger 
objects, the undigested core of each member of the sequence, 
typically contributes a density $\rho_0$ at a scale $r_c$,
satisfying the the relation $\rho_0 = c_1 r_c^{-(9+3n)/(5+n)}$,
with some constant $c_1$.

This suggests that the inner density profile of the 
bigger halo, which is the envelope of the profiles of the 
nested sequence of smaller mass cores could have the form 
\begin{equation}
\rho(r)  \propto r^{-\alpha}, \qquad 
\alpha = \alpha_n = {9 + 3n \over 5 + n}
\label{dencor}
\end{equation}
Note that for any $n < 1$, $\alpha < 2$. So one
expects the core density profile to have a power law
dependence, shallower than $r^{-2}$,
when smaller mass cores remain undigested in the formation
of a larger mass. 
It is intriguing that the same form for the {\it density profile}
(as against the correlation function) is also
argued for by Peebles (1980; section 26.).
In a paper which appeared during the course
of this work Syer $\&$ White (1998) give a similar
argument, for the case when bigger halos form by purely mergers of
smaller halos. Our argument (concluding with equation (\ref{dencor}))
of course neglects both previous generation of
undigested cores and secondary infall
and is only designed to model the innermost part of a currently
virializing object, where their effects on the energetics should be minimal.

One can also state the above argument 
in terms of the velocity dispersion or the rotation
velocity profiles. The typical
velocity dispersion of a collapsed halo $\sigma \propto (M/r_c)^{1/2}$.
Since the scaling argument gives $r_c \propto M^{(n+5)/6}$, we therefore 
have $\sigma^2 \propto r_c^{(1-n)/(5+n)} $. So for any $n<1$, 
smaller mass objects have a smaller velocity dispersion
and higher phase space densities than larger mass objects. 
The survival of a nested sequence of
cores during the inhomogeneous collapse to form bigger and
bigger objects, then suggests that the 
velocity dispersion profile in the core regions will
scale as 
$\sigma^2(r) \propto r^{(1-n)/(5+n)}$. For any $n < 1$, 
an alternate signature of undigested cores is then
a velocity dispersion which decreases with decreasing radius
in the above fashion. It is interesting to note in this context
that, the cluster scale halo core in the Moore {\it et al.} (1998)
simulation, does indeed show  such a velocity dispersion profile,
with $\sigma$ decreasing with decreasing $r$ 
(Moore, private communication). 

We have assumed above that each stage of the hierarchy 
arises from a typical density fluctuation. 
In general there would be a scatter in the sub-halo
properties since the initial density peaks heights are 
random with a Gaussian probability
distribution. This will lead to a scatter in the slope $\alpha$,
for any individual halo (cf. Nusser and Sheth 1999 and section 5 below).

The arguments so far have been semi-quantitative but general. We 
consider, in the next section, 
another approach to the density profiles of halo cores via
spherically symmetric, self similar collapse solutions to the 
Vlasov equation. Our motivation is to see
whether the results derived 
above, can be recovered in any simple, analytically tractable model.
This model will also allow
us to examine, in a simple setting, the dynamical
constraints on obtaining core density profiles of the form
given by Eq. (\ref{dencor}).

\section{ Self similar collapse and halo density profiles: a fluid 
approach}

Consider the collapse of a single spherically
symmetric density perturbation, in a flat background universe.
Suppose the initial density fluctuation is a power law in radius.
Then there is no special scale in the
problem either from initial conditions or cosmology. 
We expect to be able to describe the further evolution of
such a density perturbation, through a self similar solution.
FG and B85 looked at purely radial self similar collapse by 
solving for the self similar particle trajectory.
We adopt a different approach here, examining 
directly the evolution of the distribution function of the
dark matter. During the course
of this work we learned that a number
of authors (Padmanabhan 1994, unpublished notes; Padmanabhan 1996a,
Chieze, Teyssier $\&$ Alimi 1997, Henriksen $\&$ Widrow 1997)
have also adopted this approach to the self
similar collapse problem  considered by FG and B85. We will
emphasize here also the role of non-radial motions
in self similar collapse solutions.

\subsection{ The self similar solution}

The evolution of dark matter phase space 
density $f({\bf r}, {\bf v}, t)$ is governed by the Vlasov Equation,
\begin{equation}
{\partial f \over \partial t} + {\bf v}. {\partial f \over \partial{\bf r}}
+ {\bf a}. {\partial f \over \partial{\bf v}} = 0 ,
\label{Vlasov}
\end{equation}
where ${\bf r}$ and ${\bf v} = \dot {\bf r}$ are the proper co-ordinate 
and velocity of the particles respectively. Also the acceleration 
${\bf a} = \dot {\bf v} = - {\bf \nabla }\Phi$, with
\begin{equation}
{\bf \nabla}^2\Phi = 4 \pi G \rho = 4 \pi G \int f d^3 {\bf v} .
\label{pois}
\end{equation}
By direct substitution, it is easy to verify that these equations admit 
self similar solutions of the form 
\begin{equation}
f({\bf r}, {\bf v}, t) = k_2 k_1^{-3} t^{-q -2p} F( {{\bf r}\over k_1 t^p}, 
{{\bf v}\over k_1 t^q}) ; \qquad  p = q + 1 ,
\label{scalf}
\end{equation}
where $k_1,k_2$ are constants which we will fix to
convenient values below.

We have used proper co-ordinates here
since the final equilibrium halo is most simply described in these
co-ordinates. (The same solution in co-moving
co-ordinates for the density is given by Padmanabhan (1996a)). 
Defining a new set of co-ordinates 
${\bf y} = {\bf r}/(k_1t^p)$, ${\bf w} = {\bf v}/(k_1t^q)$ and a scaled
potential $\chi =k_1^{-2} t^{-2q}\Phi$, 
the scaled phase space density $F$ satisfies 
\begin{equation}
-(q + 2p) F - p {\bf y}. {\partial F \over \partial{\bf y}}
-q {\bf w}. {\partial F \over \partial{\bf w}}
+ {\bf w}. {\partial F \over \partial{\bf y}}
-{\bf \nabla}_{\bf y}\chi . {\partial F \over \partial{\bf w}} = 0 ;
\label{valsc}
\end{equation}
\begin{equation}
{\bf \nabla}_{\bf y}^2\chi = 4 \pi G k_2  \int F d^3 {\bf w} .
\label{potsc}
\end{equation}

Consider the evolution of a spherically symmetric density perturbation,
in a flat universe whose scale factor $a(t) \propto t^{2/3}$. 
For self similar evolution, the density is given by
\begin{equation}
\rho(r,t) = \int f d^3{\bf v} =
k_2 t^{2q -2p} \int F(y, {\bf w}) d^3{\bf w}
= k_2 t^{-2}\int F(y, {\bf w}) d^3{\bf w} \equiv k_2 t^{-2} \psi(y)
\label{densc}
\end{equation}
where we have defined $r = \vert {\bf r} \vert$, $y = \vert {\bf y} \vert$ 
and used the relation $p = q+1$. For the flat universe, 
the background matter density evolves as
$\rho_b(t) = 1/(6 \pi G t^2)$. So the density contrast 
$\rho(r,t)/\rho_b(t) = \psi(y)$, where we take $k_2 = 1/(6\pi G)$.

\subsection{ Linear and non-linear limits}

Let the initial excess density contrast averaged over a 
sphere of co-moving radius $x= r/a(t) \propto rt^{-2/3}$ be a power law 
$\bar\delta(x,t_i) \propto x^{-3\epsilon}$. 
Since $\rho/\rho_b$ is a function of $y$ alone, the $\bar\delta(x,t)$
will also be a function only of $y$.
Note that, in the linear regime, it is the excess density contrast
averaged over a {\it co-moving} sphere,
which grows as the scale factor $a(t)$. So one can write
for the linear evolution of the spherical perturbation
\begin{equation}
\bar\delta(r,t)= \bar\delta_0 x^{-3\epsilon}t^{2/3}= \bar\delta_0 r^{-3\epsilon}t^{2/3 + 2\epsilon} =
\bar\delta_0 y^{-3\epsilon}t^{- 3\epsilon p + 2/3 + 2\epsilon} ,
\label{lincon}
\end{equation}
where we have substituted  $r = y t^p$.
This can be a function of $y$ alone, 
for a range of $t$ in the linear regime iff
$- 3\epsilon p + 2/3 + 2\epsilon = 0$, which gives
\begin{equation}
p = {2 + 6\epsilon \over 9\epsilon} .
\label{adet}
\end{equation}
We see that once the initial density profile is specified, the 
exponents $p,q$ of the self similar solution are completely determined.
(For an initial $\bar\delta(x,t_i) \propto x^{-3\epsilon}$, the
radius of the shell turning around at time $t$ is, $r_t(t) \propto t^p$.
So a natural way of fixing the constant $k_1$ is by taking 
$k_1t^p = r_t(t)$, and $y = r/r_t(t)$. We will do this in
what follows.)

Consider now what happens in the non-linear limit.
The zeroth moment of the Vlasov equation gives 
\begin{equation}
{\partial \rho \over \partial t} + {\bf \nabla}_{\bf r}.(\rho \bar{\bf v}) = 0
\label{contm}
\end{equation}
Here $\bar{\bf v}$ is the mean velocity 
(first moment of $f$ over the velocity).
In the non-linear regime, one expects, 
each shell of dark matter, which was initially expanding, 
to turn around, after reaching a maximum radius and 
collapse. Subsequently the shell would oscillate between a minimum
radius,  which depends on how much non-radial velocities the shell particles
have and a maximum radius, which depends on how the mass within the shell
grows with time. In regions which have had a large amount of 
shell crossings, the halo particles
have settled to nearly zero average infall velocity, that is $ \bar{v_r} = 0$.
(They could of course still have velocity dispersions). Using 
$\bar{ v_r} \equiv 0$ in (\ref{contm}) , we have $(\partial \rho /
 \partial t) = 0$, in the non-linear regime. In this regime therefore, 
\begin{equation}
 \rho(r,t) = Q(r) = Q(yt^{p}) = {1 \over 6 \pi G t^{2}} \psi(y)
\label{nonc}
\end{equation}
This functional equation has only power law solution,
because of the power law dependences on $t$. 
Substituting $Q(r) = q_0 r^{-\alpha}$ into Eq. (\ref{nonc}), 
and using $r \propto yt^p$,
we obtain $y^{-\alpha} t^{-p \alpha} \propto t^{-2} D(y)$. This can
only be satisfied for range of $t$ in the non-linear regime
provided $p\alpha = 2$. So, for an initial density profile 
with a power law slope $3\epsilon$, the power law slope of the
density in the non-linear regime is given by,
\begin{equation}
\alpha = {2 \over p} = {9\epsilon \over 3\epsilon + 1} .
\label{nonpow}
\end{equation}

B85 considered the self-similar secondary infall onto an initially
localised, overdense perturbation,
corresponding to taking  $\epsilon = 1$. Using Eq. (\ref{nonpow})
this gives $\alpha = 9/4$, the slope for the density
at the non-linear end deduced by
following the self similar particle trajectory. 
FG considered a range of $\epsilon$ and our value
of $\alpha$ agrees with that obtained by them, 
again by following
particle trajectories. Both these authors also restricted 
themselves to purely radial orbits. In this case FG
argued that while the above form for $\alpha$ should obtain
for $2/3 \leq \epsilon < 1$, for $\epsilon < 2/3$, one goes to
the limiting value $\alpha = 2$. However, this is only true for
purely radial trajectories (cf. White and Zaritsky 1992;
Sikvie, Tkachev $\&$ Wang 1997). We will also see below, 
by considering the higher moments of the Vlasov equation, that
$\alpha < 2$ can only obtain if the system has 
non-radial velocity dispersions. 

What should we choose for the value of $\epsilon$? For a power 
law $P(k) \propto k^n$,
the fractional density contrast averaged over a co-moving sphere of
radius $x$, is distributed as a Gaussian, with a variance
$ \propto x^{-(3+n)/2}$ (cf. Peebles 1980).
This suggests  a "typical" spherically averaged initial
density law for a halo collapsing around a randomly placed point
of the form $\bar\delta(x,t_i) \propto x^{-(3+n)/2}$, or
$3\epsilon = (3 + n)/2$. Suppose we use this value of $\epsilon$ for
the initial density profile of a halo. Then in the non-linear stage,
the halo density in regions which have settled down to a zero mean
radial velocity, will be $\rho(r,t) \propto r^{-\alpha}$, where, 
using $ 3\epsilon = (3 + n)/2$ in Eq. ( \ref{nonpow} ) 
\begin{equation}
\alpha = \alpha_n = { 9 + 3n \over 5 + n}
\label{aln}
\end{equation}
This result should obtain for collapses from power law
initial power spectrum.
Remarkably, this is the same law we derived 
earlier for the core of a collapsed halo, assuming that the cores of
sequence of sub-halos are left undigested, during the formation
of the bigger halo. 

An alternate choice, $3\epsilon = (3+n)$ would be relevant
if one were considering the collapse around an isolated
high density peak; since in this case the initial density 
profile would be proportional to the correlation function 
to lowest order (cf. Bardeen {\it et al.} 1986).
In this case one gets $\alpha = (9+3n)/(4+n)$
(Hofmann $\&$ Shaham 1985, Padmanabhan 1996b).
(Since $\epsilon < 1$ for overdense  perturbations, we can use
this choise only for $n < 0$).

Note that for $n < 1$ the density law given by (\ref{aln}) 
is shallower than $1/r^2$,
which was claimed to be a limiting form by FG in case of 
radial collapse. To see how such a restriction
comes about and when one can obtain 
a shallower slope than $r^{-2}$ for the halo cores, 
it is interesting to consider
the higher moments of the Vlasov equation (the Jeans equations) for
the spherical self similar solution. 

\subsection{A fluid approach to collisionless dynamics }

Suppose we multiply the Vlasov equation by the components of 
${\bf v}$ and integrate over all ${\bf v}$. 
In regions where large amounts
of shell crossing have occurred, one can assume that 
a quasi "equilibrium" state obtains,
whereby all odd moments of the distribution function, over 
$({\bf v} - \bar{\bf v})$, may be neglected. 
Assume there is no mean rotation to the halo, that is
$\bar v_{\theta} = 0$ and $\bar v_{\phi} = 0$. Then we get
\begin{equation}
{\partial(\rho \bar v_r) \over \partial t} 
+{\partial(\rho \bar{v_r^2}) \over \partial r} 
+{\rho \over r} (2\bar{v_r^2} - \bar{v_{\theta}^2} - \bar{v_{\phi}^2})
+ {GM(r)\rho \over r^2} = 0 ;
\label{radm}
\end{equation}
\begin{equation}
\bar{v_{\theta}^2}  = \bar{v_{\phi}^2} .
\label{thetm}
\end{equation}
Here $M(r)$ is the mass contained in a sphere of radius $r$.

For a purely radial collapse we can set 
$\bar{v_{\theta}^2}  = \bar{v_{\phi}^2} =0$.
Let us also assume to begin with that one
can set $\bar v_r = 0$, in the inner parts. 
Then integrating the Jeans equation 
( \ref{radm} ),  with $\rho = q_0 r^{-\alpha}$ gives 
\begin{equation}
\bar{v_r^2} = r^{2- \alpha} \left [{4\pi G q_0 \over 
2(\alpha -2 )(3-\alpha)} \right ]
\equiv {1 \over (\alpha -2 )} {GM(r)\over 2r} . 
\label{consisr}
\end{equation}
So purely radial self-similar collapse with no tangential
velocities, and with $\alpha > 2$, leads to a
radial velocity dispersion in the core which scales as
$\bar{v_r^2}\propto r^{-(\alpha - 2)}$.
This agrees with the
radial velocity dispersion scaling as $r^{-1/8}$ for B85
gaseous collapse solution. ($\alpha = 2$ needs to be treated separately).
Further the RHS of Eq. (\ref{consisr}) should be necessarily non-negative,
which is violated when $\alpha < 2$. If one has a purely
spherically symmetric collapse and zero tangential
velocities, then the density law cannot become shallower
than $\alpha=2$ and maintain a static core with 
$\bar{v_r}=0$.  This agrees with FG. 
Our example illustrates a point we mentioned earlier.
Even if simple scaling arguments suggest a $\alpha < 2$
possibility, there could be dynamical restrictions
for realizing such core profiles.

Let us now include the effect of tangential velocity dispersions. 
The Jeans equation gives 
two equations for the three unknown velocity
dispersions, even for a static core. 
To see if one can close the system let us look
at the second moments of the Vlasov equation (the energy equations)
We get
\begin{equation}
{\partial(\rho \bar{v_{\theta}^2}) \over \partial t} 
+{1 \over r^4}{\partial(\rho r^4<v_rv_{\theta}^2>) \over \partial r} -
{2\rho <v_{\theta}v_{\phi}^2> cot\theta  \over r}
 + {\rho \bar{v_{\theta}^3} cot\theta \over r} = 0 ,
\label{thetsqm}
\end{equation}
\begin{equation}
{\partial(\rho \bar{v_{\phi}^2}) \over \partial t} 
+{1 \over r^4}{\partial(\rho r^4 < v_rv_{\phi}^2 >) \over \partial r} 
+ {\rho < v_{\theta}v_{\phi}^2 > cot\theta  \over r} = 0 ,
\label{phisqm}
\end{equation}
\begin{equation}
{\partial(\rho \bar{v_r^2}) \over \partial t} 
+{1 \over r^2}{\partial(\rho r^2\bar{v_r^3}) \over \partial r} -
{2\rho < v_r(v_{\theta}^2+v_{\phi}^2) >  \over r} 
+2 \bar{v_r} \rho {GM/r^2} = 0 ,
\label{radsqm}
\end{equation}
where $M = \int 4 \pi r^2 \rho$ is the mass within $r$
and both $<>$ or a bar over a variable denotes
a normalized moment over $f$.

Consistent with our statistical assumption for the core regions, we
assume that initially the tangential velocities have zero skewness.
Then in purely spherically symmetric
evolution they would not develop any skewness, that is 
$\bar{v_{\theta}^3} = \bar{v_{\phi}^3} =
< v_{\theta}v_{\phi}^2 > = 0$ for all times.
Also if the initial velocity ellipsoid had one
of its principle axis pointing radially, we do not expect this axis
to become misaligned in purely spherical evolution.
This means we can assume $< v_r v_{\theta}^2 > =
\bar{v_r} \bar{v_{\theta}^2 } = 0 $ in the static core.
Eq. ( \ref{thetsqm} ) then implies 
$(\partial(\rho \bar{v_{\theta}^2})/\partial t) = 0$
or $\rho \bar{v_{\theta}^2} = K(r)$ independent of $t$.
For the scaling solution we then have
\begin{equation}
\rho \bar{v_{\theta}^2} = K(r) = K(yt^p) = k_2k_1^2
t^{4q -2p}\int w_{\theta}^2 F(y,{\bf w})d^3{\bf w} .
\label{tan}
\end{equation}
Once again substituting a power law solution $K(r) = K_0 r^s$,
to this functional equation, we get the constraint from matching
power of $t$ on both sides,
$ps =4q - 2p$. Using $p = q +1$,
we then get
$s = 2 - 4/p = 2 - 2\alpha$, and so
\begin{equation}
\rho \bar{v_{\theta}^2} = K_0 r^{2 - 2\alpha} .
\label{tanvel}
\end{equation}

Integrating the radial momentum equation using
Eq. (\ref{radm}) , (\ref{thetm}), (\ref{tanvel}) and using 
$\rho = q_0 r^{-\alpha}$. Equation (\ref{consisr}) for the radial 
velocity dispersion is now altered to
\begin{equation}
\bar{v_r^2} = r^{2 - \alpha} \left [ {K_0 \over (2 - \alpha) q_0} - 
{4\pi G q_0 \over 2(2-\alpha)(3-\alpha)} \right ]
\equiv {1 \over (2 - \alpha)} \left [ \bar{v_{\theta}^2}(r) 
- {GM(r)\over 2r} \right ] .
\label{consist}
\end{equation}

Several important points are to be noted from the
above equation. A crucial one is that, when $ \alpha < 2$, 
the RHS of  Eq. (\ref{consist}) can 
remain positive, only provided that one has a non zero tangential
velocity dispersion. In fact, for any $\alpha < 2$, one
needs the tangential velocity dispersion to be at least
as large as $GM/2r$, comparable to the gravitational potential
energy per unit mass.
Also one can see that to obtain static cores with $\alpha < 1$, 
the required tangential dispersion must be 
larger than the radial velocity dispersion. So if 
in halo cores tangential velocity dispersions are constrained
to be smaller than radial velocity dispersions, then a
core density profile shallower than $1/r$ cannot obtain
in the self-similar case.
Also note that for $\alpha < 2$, all the components
of velocity dispersions decrease with decreasing radius,
as suggested by the simple scaling arguments of the previous section.

In a realistic collapsing halo it is quite likely that particles
develop non-radial velocities. Tidal forces by mass concentrations
outside the halo and the presence of substructure within the collapsing
halo will lead to non-radial motion of particles. 
More generally the process of violent relaxation during the inhomogeneous
collapse to form the halo will lead to a more isotropic velocity
dispersion. 

The above results for the halo core arise simply
from the properties of the self similar solution 
and the assumption of a static core. From the energy equation
(\ref{radsqm}) we note that a time independent
radial velocity dispersion, can only obtain 
if the radial velocity skewness $<(v_r -\bar{v_r})^3>$
is also zero. Note that in the core regions where large amounts
of shell crossing has occurred, as we stated earlier, the
radial skewness is indeed expected to be small. So for
the core regions one can in fact make this statistical assumption.
Such a treatment will correspond to considering a fluid like 
limit to the Vlasov equation. 

However, the radial skewness
will become important near the radius, where infalling matter
meets the outermost re-expanding shell of matter. This region
will appear like a shock front in the fluid limit.
A possible treatment of the full problem in the fluid approach 
to the Vlasov equation then suggests itself. This is
to take the radial skewness to be zero both inside and outside a
"shock or caustic" radius, whose location is to be
determined as an eigenvalue, so as to match the inner core
solution that we determine in this section
with an outer spherical infall solution. 
One has to also match various quantities
across this "shock", using jump conditions,
derived from the equations themselves.
To do this requires numerical solution of the self consistent
set of moment equations, to the scaled Vlasov equation.
The details of such a treatment are given a
companion paper (Subramanian 1999, S99). 
Here we summarise the general conclusions of this work.

The numerical results in S99 shows the importance of
tangential velocity dispersions, in deciding whether the
self similar solution, with an initial density profile
shallower than $1/r^2$ ($\epsilon < 2/3$) retains a memory of this initial
profile or whether the density profile tends to a universal $1/r^2$ form.
The set of solutions show that for
a large enough ${\bar v_{\theta}^2}/{\bar v_r^2} > 1$, the
the core density profile is indeed close to the form 
$\rho \propto r^{-\alpha}$,
with $\alpha = 9\epsilon/(1+3\epsilon)$. 
For ${\bar v_{\theta}^2}/{\bar v_r^2} \sim 1$, 
some memory of the
initial density profile is always retained; the density profile
has an asymptotic form $\rho \propto r^{-\bar\alpha}$, with
$ \alpha < \bar\alpha < 2$. 
When ${\bar v_{\theta}^2}/{\bar v_r^2} << 1$, the density profile goes
over to the $1/r^2$ form derived by FG. Also for
very shallow initial density profiles with $\alpha  < 1$,
one must necessarily have a tangential dispersion much larger
than radial dispersion to get a static core region, 
retaining the memory of the initial density profile.

The spherical self similar collapse solutions provide a
useful means of examining the dynamics of dark halo formation, and
its implications for the core-density profiles,
although limited by the spherical symmetry assumption.
A complimentary approach would be direct cosmological
N- Body simulations of halo formation, which we 
will consider in Section 5. Before this, we consider
briefly below, the outer profiles of dark halos, especially in
low density universes.

\section{ Outer profiles of dark halos}

In the cosmological context, as mentioned in section 2, 
any collapsed mass will continue to accrete surrounding material,
as long as the matter density dominates the energy density. 
We now analyze the consequence of such 
secondary infall for the outer profile of halos, 
following the Gunn-Gott paradigm, relaxing the
restriction of a flat universe. 

Consider the collapse a spherically symmetric density perturbation,
in a universe with present matter density parameter $\Omega_m$ 
and cosmological constant $\Lambda$. Let the initial
density distribution (at time $t_i$), be $\rho(r,t_i) = \rho_b(t_i)
(1 + \delta_i(r))$, and initial velocity 
of the perturbation be the Hubble velocity.
Here $\rho_b(t)$ is the matter density of the smooth universe,
and $\delta_i$ the initial fractional density contrast of the
perturbation, as before.
Consider a spherical shell initially at a radius $r_i$. 
The evolution of the proper radius $r(t)$
of any such shell before shell crossing 
is governed by 
\begin{equation}
{1 \over 2}\left({dr\over dt}\right)^2 - {GM\over r} 
-{\Lambda r^2 \over 6} = E(M) .
\label{energ}
\end{equation}
Here $M = \rho_b(t_i) (4\pi r_i^3/3)(1 + \bar\delta_i)$,
is the mass enclosed by the shell and 
\begin{equation}
\bar\delta_i(r_i) = {3 \over r_i^3} \int_0^{r_i} \delta_i(u) u^2 du
\label{avd}
\end{equation}
is the spherically averaged value of $\delta_i(r)$ within $r_i$.
The "energy" $E(M)$ can be fixed by evaluating the LHS of
Eq. (\ref{energ}) at the initial time. The shell
will turn around at a time say $t_m$, when $dr/dt = 0$,
and its radius is say $r(t_m,r_i) \equiv R(r_i)$.
Setting $dr/dt =0$ in Eq. (\ref{energ}) gives 
\begin{equation}
{(1 + \bar\delta_i) \over y} + y^2 \lambda 
= \lambda  -(\Omega_{mi}^{-1} - 1) +\bar\delta_i 
=\bar\delta_i + {(\Omega_t - 1) \over \Omega_{mi}} ,  
\label{turn}
\end{equation}
where the total value of $\Omega$ (including the 
cosmological constant) is $\Omega_t = \Omega_{mi}(1 + \lambda)$.
Here $y \equiv R(r_i)/r_i$ and
$\lambda = \Lambda /(3 H_i^2 \Omega_{mi})$ with
$H_i$ the Hubble parameter and $\Omega_{mi}$ the
matter density parameter at the initial time $t_i$.

Let us begin with the case $\Lambda = 0$.
Then Eq. (\ref{turn}) gives
\begin{equation}
y \equiv {R(r_i) \over r_i} = {(1 +\bar\delta_i) 
\over \bar\delta_i - \delta_c} ,
\label{turopen}
\end{equation}
where we define $\delta_c \equiv (\Omega_{mi}^{-1} - 1)$.
In an open universe with $\Omega_{mi} < 1$,
one needs $\bar\delta_i > \delta_c$ for a shell to turn around
and collapse. For a monotonically decreasing initial density profile,
there will then be an outer most shell with an initial
radius $r_c$, satisfying $\bar\delta_i(r_c) = \delta_c$, 
such that only shells with $r_i < r_c$, can recollapse onto the
density perturbation. To work out the outer profile after collapse,
we need to know the final effective radius $r(R)$, of a shell 
turning around at radius $R$. Following the Gunn-Gott picture,
we assume that these two radii can be related by
$r(R) = f_0R$, where $f_0$ is some constant. Note that
$f_0$ is indeed a constant if the collapse is exactly self similar,
and the initial density profile is sufficiently steep
( Section 3, with $\epsilon > 2/3$). 
For deriving the dominant scaling of the outer
profile with radius, even in an open universe, 
it should also suffice to treat it to
be approximately constant (see below).
Consider an initial density profile
with $\bar\delta_i(r_i) = d_0 (r_i/r_0)^{-3\epsilon} $.
Then using mass conservation, $4\pi r^2 dr \rho(r) = 
4\pi r_i^2 dr_i \rho_b(t_i) (1 +\delta_i)$, and 
the final density profile $\rho(r)$
after collapse is given by
\begin{equation}
\rho(r) = {\rho_b(t_i) \over f_0^3} { \bar\delta_i^3(r_i)
\left[1 - (r_i/r_c)^{3\epsilon}\right]^4 \over
\left[1 + 3\epsilon - (r_i/r_c)^{3\epsilon} \right] }
\label{rhof}
\end{equation}
with
\begin{equation}
r_i =  r { \bar\delta_i [1 - (r_i/r_c)^{3\epsilon} ] \over f_0}
\label{rrirel}
\end{equation}
(Here we have also assumed that the initial $ \delta_i << 1$.)

Two limits are of interest. First, for a flat matter 
dominated universe we 
have $\delta_c = 0$ and $r_c \to \infty$. Then 
$[1 - (r_i/r_c)^{3\epsilon} ] \to 1$ and we recover
the standard result  
\begin{equation}
\rho(r) = {\rho_b(t_i) \over 1 + 3\epsilon } 
\left({d_0 \over f_0} \right)^{3/( 1+3\epsilon)} 
\left({r \over r_0} \right)^{- 9\epsilon /(1 + 3\epsilon)}.
\label{rhofl}
\end{equation}
For $\epsilon = 1$, which is the steepest possible value
obtaining for an initially localised overdense perturbation,
we recover a halo profile $\rho(r) \propto r^{-9/4}$ (B85 and Section 3).

Now we return to the case of an open universe but examine the
outermost profile where $r_i \to r_c$, 
the critical radius. In this case, from (\ref{rhof}) and 
(\ref{rrirel}), the outer profile at large radii is given by
\begin{equation}
\rho(r) \to {\rho_b(t_i) f_0 \over 3\epsilon \delta_c} 
\left({r \over r_c} \right)^{- 4},
\label{rhoflo}
\end{equation}
the slope being independent of the initial power law
slope $\epsilon$.
This interesting result seems to have been already known to Gunn (1977),
but not much emphasised since then.
As the outer profile in this case is
also a pure power law, it is likely that our assumption 
of a constant $f_0$ is valid in this case. It would be of
interest to find an exact similarity solution of the
form given in Section 3, and valid in an open universe
which recovers this outer profile.

Let us now turn to the currently
popular flat cosmological models with $\Lambda \ne 0$. 
First from Einstein's equation for the scale factor,
we have $\lambda - \Omega_{mi}^{-1} + 1 = 0$ 
(or $\Omega_t = 1$). Using this 
and taking $\bar\delta_i << 1$, 
Eq. (\ref{turn}) for $R(r_i)$ becomes
\begin{equation}
{1 \over y} + y^2 \lambda  = \bar\delta_i .
\label{turnc}
\end{equation}
For a monotonically decreasing $\bar\delta_i(r_i)$,
the RHS of (\ref{turnc}) monotonically decreases.
However the LHS as a function of $y$ has a 
minimum value at $y =y_c = (2\lambda)^{-1/3}$,
which is given by $\delta_{\lambda} = (3/2) (2\lambda)^{1/3}$. 
There again exists a critical radius $r_\lambda$
defined by $\bar\delta_i(r_\lambda) = \delta_\lambda$,
such that, only those shells with $r_i < r_\lambda$ will be able to
turn around and collapse. For shells with 
initial radii $r_i > r_{\lambda}$,
the repulsion due to the cosmological
constant overcomes the attractive gravitational force,
and so they expand for ever.
(The critical value of $y=y_c$ can also be written
as $y_c = 3/(2\delta_\lambda)$, a limit got by
Barrow and Saich (1993; eq. 26)).
Although this feature is similar to the open universe
case, there is a major difference between the
open model and the flat universe with a
cosmological constant. In the $\Lambda$ dominated model,
even for the limiting case $r_i \to r_\lambda$
the turn around radius tends to a finite limit,
$R(r_i) \to r_\lambda y_c$. In the open model
on the other hand, as $r_i \to r_c$,
(the limiting critical radius beyond which shells expand for ever), 
$\bar\delta_i \to \delta_c$,
and $R(r_i)\propto (\bar\delta_i -\delta_c)^{-1} \to \infty$.
But in both cases the accreted mass is finite and equal to that 
initially within $r_c$ or $r_{\lambda}$.

Let us now consider the limiting denity profile about this
outer most cut off radius, for the $\Lambda$ dominated model.
For this, it is sufficient to consider values of $r_i$ 
close to but less than $r_\lambda$ and expand the LHS of Eq. (\ref{turnc}),
about $y = y_c$, retaining upto quadratic terms in
$(y -y_c)$. We then have for the outermost shells 
$\delta_\lambda + 3\lambda (y -y_c)^2 = \bar\delta_i(r_i)$ or
\begin{equation}
R(r_i; r_i \to r_\lambda) = r_i y_c - (3\lambda)^{-1/2} 
\left[ \bar\delta_i(r_i) -\delta_\lambda \right]^{1/2}r_i ,
\label{turnl}
\end{equation}
where the negative square root has to be taken
as the turn around radii for the collapsing shells are smaller
than the maximum value of $R = r_\lambda y_c$.
For computing the collapsed density profile,
we need again the relation between the turn
around radius and the final effective radius.
For a $\Lambda$ dominated model, it is known that
$f_0$, the ratio of the effective "virial" radius to
the turn around radius,
depends on the turn around radius itself, 
that is $r = f_0(R) R$ (Lahav {\it et al.} 1991; Barrow and Saich 1993). 
So $dr/dr_i = f_0 (dR/dr_i) (1 + c_1)$, 
where $c_1 = d(ln(f_0)/d(ln(R)$. Let us assume
the power law form of $\bar\delta_i(r_i)$ given above.
From mass conservation once again $\rho(r) = 
\rho_b(t_i) (r_i/R)^2 (dr_i/dR) f_0^{-3} (1 + c_1)^{-1}$.
From  Eq. (\ref{turnl}), in the limit of $r_i \to r_\lambda$,
which is the limit relevant for evaluating the outer profile,
$(r_i/R) \to y_c^{-1}$ a finite value. However
from this equation $dr_i/dR \to 0$ and the outer density
profile cuts off as one nears a critical radius.
Using Eq. (\ref{turnl})
the limiting outer halo density profile becomes
\begin{equation}
\rho(r) \to  {4 \lambda \rho_b(t_i) \over 3 \epsilon f_0^3 (1 + c_1)}
\left [ 1 - (r/\bar r_\lambda)^{3\epsilon} \right]^{1/2} ,
\label{cosmo}
\end{equation}
where $\bar r_\lambda = r_\lambda /(y_c f_0)$.
(Here we can treat $f_0$ and $c_1$ as constants 
evaluated at the limiting $R = \delta_\lambda y_c$).
The above profile shows that in a universe dominated
by a cosmological constant, the mass of halos
is again convergent. We caution that the above forms
for the outer profile of halos in open models (viz. Eq. (\ref{rhoflo}) and 
(\ref{cosmo}) ), obtain only near the 
cutoff radius and only at late times, as the outermost bound shell
turns around only as $t \to \infty$. 
For a finite fixed time $t$ and general $r_i$ a more detailed 
solution of (\ref{energ}) and (\ref{turnc}) is needed, for
finding the density profile. Of course in the innermost regions,
the $\lambda y^2$ term in (\ref{turnc}) is
expected to be small compared to $1/y$ term,
and we will recover the results of the 
flat model, discussed above and in Section 3.
We now turn to the study of halo
properties through direct cosmological 
N-body simulations.

\section{ Halo properties through cosmological N- body simulations}

In order to clarify the effect of non-linear evolution
in determining the structure of dark halo cores,
it is best to look at power-law spectra, as this has
no special scale, rather than a model like the CDM.
We have therefore simulated three power law models with index
$n=-2,-1,0$, with $\Omega_0=1$.  Each simulation is run
using a particle-mesh (PM) code,
with $768^3$ mesh points and $256^3$ particles.
Although each model has no intrinsic scale,
we choose the box size of each simulation to
be comoving $10h^{-1}$Mpc. At the end of the simulation, 
which we identify with redshift zero, the 
rms density fluctuations on a $8h^{-1}$Mpc sphere, 
$\sigma_8 = 1/1.3$ in the $n=-2$ model, while in the 
$n=-1$ model, $\sigma_8 = 1/1.5$ and in the $n=0$ model,
$\sigma_8 = 1/4$. Thus, in all cases waves larger than
the box size have typically not grown to the non-linear
domain. We will use this scale notation to facilitate 
discussions at the relevant scales.
The nominal spatial resolution of the simulations 
is $13h^{-1}$kpc, adequate for resolving the galaxy size halos
that we are interested in here.
The mass of each particle is $1.65\times 10^7h^{-1}$M$_\odot$.
Thus a galaxy size halo would contain of order $10^5$ particles,
essential for our purpose of computing
density and velocity dispersion profiles. The highest mass
resolution in the present simulations is also required to avoid 
two-body relaxation in the inner regions of halos, which we are
most interested in. Within the innermost bin of our calculation
($10h^{-1}$kpc, see below) at an overdensity of about $10^5$ 
(see Figures (2,3,4) below) there are of order thousand 
particles, thus two-body relaxation
should be negligibly small (see also Eq. (\ref{trel}) below).

In each simulation, the center of each halo 
is selected as the local maximum of the mass distribution  within
spheres of comoving radius of $10\rm h^{-1}$kpc.
The density profile of each halo is calculated
using spherical averaging with a logarithmic bin size of 0.02 dex.
The velocity dispersions (both tangential and radial) are computed 
in the restframe of each spherical shell.
We have used the density and velocity dispersion profiles of 
20 halos in each model for the analysis to be described below.
In particular we would like to examine if the halo density profiles
show  evidence for the scaling laws of Section 2 and 3, and a dependence 
on the power spectrum.

In our analysis of the halo density profiles, for each halo
we first fitted $\rho(r)$, by a double power law model of the form given by
\begin{equation}
\rho(r) =  \left ({r_c \over r}\right )^{\alpha_0} 
{ \rho_0 2^{\beta_0} \over [1 +(r/r_c)]^{\beta_0}}.
\label{den}
\end{equation}
We used the log density versus log radius data, taking
all radii within a nominal "virial radius", say $r_v$, where the density
dropped to about $200$ times the background density. 
We made these fits by using an IDL routine (curvefit) which 
takes a trial model set of parameters and iterates them
to minimize the squared sum of the deviations of the fit from
the actual density profile. By judicious choice of the
starting values of the parameters, it is relatively
simple to obtain good convergence in the set of model parameters.
The density profile of the 20 halos
in each model $n=0$, $n=-1$, and $n=-2$ are shown as dotted
lines in Figures 2, 3 and 4 respectively.
(Here the density has been normalised with respect to
$\rho_c = \rho_b(t_0)$, the background density at the present epoch $t_0$).
The converged model fit is also shown as a light solid line in these figures.
One can see from the figures that, in general, the
double power law fit is excellent. Infact, for every halo,
the fractional deviations, of the actual log$(\rho)$
compared to the model fit, are very small;
in general much less than $1\%$ for most radii, 
with maximum deviations less than a few percent. 
Further as in these fits one minimizes the total
least square deviations of the data from the fitted
profile, the fitted function is moderately robust to local
perturbations in the density.
We can therefore use the model fitted profiles to study 
the properties of halo density profiles. 

For each halo, we then calculated the local power law
slope of the density profile by evaluating, $s(r) = d(ln \rho)/d(ln r)$,
from the model fit. We plot this local slope, $s(r)$, for every halo, 
as a thick solid line in the same plot as the density profile plot, 
in Figures 2, 3 and 4.
These $s(r)$ plots give the most detailed information
regarding the slope of the halo density profiles.
If the density profile has a power law regime,
then for this radius range, $s(r)$ would be a straight horizontal line.
Some general conclusions can be drawn from the figures themselves.
First for almost all the halos,
$s(r)$ keeps increasing monotonically with radius; showing
that the halo density profiles are in general curved, and
that the density profile keeps steepening with radius.
Previous workers like NFW have in general adopted model
double power law profiles to fit halo density profiles,
with specific values of the inner slope $\alpha_0 = 1$ and
outer slope $\beta_0 + \alpha_0 = 3$. We find
the hetrogeneity of the density profiles to be striking.
No simple formulae with fixed ($\alpha_0$,$\beta_0$) can fit this data.
Indeed for an unbiased double power law fit, the innermost value
of $s(r)$ lies between $1 - 2 $, shows a general increase with increasing $n$,
and is in general not equal to $1$. Also, 
the outermost value is generally not equal to $3$.

In order to quantify these conclusions better,
and since we have a moderately large (20) number of halos in each
model, we can infact look at the the statistics of
the inner core and outer slopes, for each model of structure
formation (given by the power spectrum index $n$).
We do this by looking at the distribution of $s_i(r_0)$
and $s_i(r_v)$ for the 20 halos in each model with a given
value of $n$. Here $s_i(r)$ is the slope function of 
the $i$'th halo in a given model, calculated from the model fit, 
and $r_0$ is some fiducial characteristic inner radius of a halo.

In Figure 5, we have given a histogram of the
distribution of the inner core slopes $s(r_0)$, for the different
models of structure formation with $n=0, -1$ and $-2$,
adopting three different values for $r_0$. In the
left hand side of this plot, we have taken $r_0$
to be the innermost radius $r_i$ of the halo,
in the middle histograms we have taken it 
to be a fixed percentage ($10\%$, say) of the virial radius,
with $r_0 = 0.1 r_v$,
while in the right most histograms have taken 
a larger value $r_0 = 0.15 r_v$.
For all the halos in the $n=-1$, or $n=0$ models,
these 3 choices correspond to progressively larger and larger 
value of $r_0$. For $n = -2$ case for about half the halos,
the innermost radius $r_i$ is of order $0.1 r_v$; 
hence the close similarity of the histograms for these two cases.
The solid arrow in each of these histogram plots shows the
location of the median value of the distribution. We also show for
comparison by a thin arrow, the location of the core-slope
$\alpha_n = 3(3+n)/(5+n)$, expected on the basis of the scaling
arguments of section 2.

From this figure we see first
that all halos do have a cuspy inner density profile,
mostly with $s(r_0) > 1$. Further,  
the core slopes show a clear spectral index dependence although 
the inner power laws are all somewhat steeper than the 
$\alpha_n = 3(3+n)/(5+n)$ form, predicted by the scaling laws. 
For example, the median value of the  core slope for the
$n=-2$ models is $s(r_i) \sim 1.3$ (leftmost histogram),
(compared to $\alpha_n = 1$), 
while for the $n = -1$ and $n= 0$ models, the corresponding
median value of the core slope shifts to $s(r_i) \sim 1.6$ 
($\alpha_n = 1.5$) and $s(r_i) = 1.8$ ($\alpha_n = 1.8$), respectively. 
For any fixed $n$, there is also a systematic increse of the median slope 
as one increases $r_0$ and goes from the left-most to 
the right most histogram. This is to be expected  as we
have a curved and continuously steepening
density profile. However the
trends between different models remain (a steeper 
inner profile for larger n).
This can be seen already for example by comparing 
the left and right side of Figure  5.

Further, we checked using the Kolmogorov-Smirnov
two-sample test whether the distribution of core slopes
$s(r_i)$ for different values of $n$, are drawn from the same
population or not. We used the one-tailed test to decide
if the vaues in one sample (say $n=0$) are stochastically
larger than the values of the other sample (say $n=-2$).
In this test one computes the 
two-sample statistic $D_{M,N} = max[S_M(X)- S_N(X)]$,
where $S_M(X), S_N(X)$ are the cumulative probability distributions
of the two samples with $M$ and $N$ number of points respectively
in each sample (in our case $M=N=20$). The
value of $NMD_{M,N}$ being larger than a given number is then
used to rule out the hypothesis (that the samples are drawn from the same
population) at various levels of confidence 
(cf. Siegel and Castellan 1988, pg. 144).
This test shows that the distribution
of core slopes of the $n=0$ model is stochastically larger than the
core slopes of the $n=-2$ model, and not drawn from the same population,
at a $ 99\%$ confidence level. The hypothesis that the core slopes of the 
$n=-1$ and $n=-2$ are drawn from the sampe population is ruled out at
a weaker $90\%$ confidence level. And the hypothesis of
the core slopes of $n=0$ and $n-1$ models being drawn from the
same population is ruled out at a $90 -95 \%$ confidence level,
depending on the binning used for the data.
Our preliminary conclusion, therefore,  
from analyzing these cosmological N- body simulations (cf. Figures 2 - 5),
is that the core density profiles of dark matter halos, do depend on 
the initial power spectrum of density fluctuations; becoming steeper
as the spectral index increases from $n = -2$ to $n =0$.

In Figure 6, we have given the corresponding
histogram for the distribution of the outer
slopes $s_i(r_v)$, for different models
of structure formation. We see from the figure that
the distribution of outer profiles is fairly
broad. For the models with $n =0$ and $n=-1$, they are
spread, with large deviations, about a median value 
$s(r_v) = 3.06$ and $s(r_v) = 3.02$ respectively. For the
halos in the $n=-2$ simulation the outer profile
is somewhat shallower, being spread around a median value
$s(r_v) = 2.55$. These results for the
outer profile suggest a large scatter about the favored NFW
value of $\beta_0 + \alpha_0 = 3$.

We have summarised the information on the core and outer slopes
of dark matter halos, 
in different models, as a scatter plot in Figure 7.
Each point in these scatter plot marks the value of the
inner core and outer slope for a particular halo. We also show as
a solid cross the location of the median value of the distributions
of slopes, with the extent of the cross giving the
$\pm \sigma_m$ error on the median. (We adopt an 
 error on the median $\sigma_m = c_N\sigma_N/\sqrt{N}$, where
$\sigma_N$ is the standard deviation of the $N$ values of the
the slope distribution and $c_N = 1.214$ 
(cf. Kendall and Stewart 1963, pg 327).
This figure further illustrates the result that the distribution of the
core and outer slopes have a large scatter but appear to display 
systematic trends as one goes from $n=0$ to $n=-2$.

At this point we should add a note of caution regarding
the determination of the inner slopes. Ideally for determining
the inner core properties one has to have a resolution as small
as possible and as many particles within the virial radius as possible,
though it is not at present clear what these numbers should be. 
In the biggest halos we have few times $10^5$ particles; and 
our resolution in these halos is about $5 - 7.5\%$ of the virial
radius. The larger resolution scale is because we have extracted 
halos from a cosmological PM simulation, though it is one of the 
best resolved PM simulation (with a $768^3$ mesh). 
Of course one advantage of the present work is that
we have a large number of halos (20) in each model,
and so can look at the halo properties in a
statistical fashion as well. And we saw above that the statistical
analysis reveals the trend in core slopes, more clearly.
At the spatial resolution of the current simulations,
the overdensity is about $10^4-10^5$, which is about the 
overdensity of real galaxy or cluster halos on the 
same scale. Therefore, our simulations may not be severely spatial
resolution limited for the present purpose of examining 
the properties of halos on these scales and larger.
Still, it would be very useful 
to see whether the trends 
we have found here for the core slope, hold
up with further high resolution
studies of a large number of halo density profiles.
In particular, it will be of great interest to go to even higher
overdensity, more inner regions of halos using higher mass and
spatial resolution simulations to further test the present 
findings from simulations as well as our analytic results.

Apart from the density profiles, it is also of
interest to study the velocity dispersion
profiles of the halos, to see if there is any evidence for
undigested cores. As we argued in section  2 and 3, this will
lead to a rising velocity dispersions in the core regions,
of the form $\sigma \propto r^{(1-n)/(2(5+n))}$, which also
implies a mild but systematic spectral dependence.
It is also of interest to check the relative importance
of tangential and radial dispersions in the halo.
Recall from the work of section 3, that tangential
dispersions are needed to get cuspy density profiles shallower
than $1/r^2$ in self similar collapse from power
law initial density profiles; and that a cuspy
profile shallower than $1/r$ required tangential
dispersions to dominate radial dispersion.
Our data on the velocity dispersion profiles are 
too noisy for drawing very firm conclusions.
However we do find most halos showing a rise in the 
radial velocity dispersion with increasing radius 
in the core regions. Further,
the tangential dispersions are smaller than radial and also
in general show much weaker (sometimes no) rise with 
increasing radius in the core.

One may wonder about the importance of 2-body relaxation effects 
in determining the properties of the halo cores in the simulation;
for example significant 2-body relaxation could lead
to an artificial steepening of the density profiles of
the halos in the core regions. We can use the halo
properties in the simulation itself to check this,
using the standard estimate for the 2-body relaxation
time scale $t_{rel}$ (cf. Binney and Tremaine, 1987 (Eq. 8.71);
see also Steinmetz and White 1997),
and comparing it with the Hubble time $t_{hub} = t_0$.
We obtain 
\begin{equation}
{t_{rel} \over t_{hub}} = 11.8 h^{-1} \left({\sigma \over \sigma_*}\right)^3 
\left({m_* \over m}\right) \left ((\rho/\rho_b) \over d_* \right )^{-1}
\left({{\rm ln}(\Lambda) \over 10} \right)^{-1} ,
\label{trel}
\end{equation}
where we have taken fiducial values $\sigma_* = 200$ km s$^{-1}$,
$m_*  = 1.65 \times 10^7 M_\odot$, $d_*  = 4 \times 10^4$,
and ln$(\Lambda)$ is the usual "coulomb" logarithm, which is
of order $10$.  In general we find this number is much larger 
than unity for all halos, even in the innermost regions. 
So 2-body relaxation is not expected to be important in the halo cores. 

\section{Discussion and Conclusions}

We have concentrated in this work on
the structure of the cores of dark halos in hierarchical clustering theories
of galaxy formation. In such theories, it is very likely that cores
of dark matter halos harbor undigested earlier generation
material. Their density structure, in physical as well as phase space,
will reflect the times and the cosmological densities when the core
material was gathered. 

In a flat universe with a power spectrum $P(k) \propto k^n$,
a consequence of undigested cores, could be a cuspy core density profile, 
shallower than that of a
singular isothermal profile, having a velocity dispersion
profile and rotation curve, which rises with increasing radius. 
Scaling arguments,
incorporating energy and mass conservation, suggests a
form for the core density profile, $\rho(r) \propto r^{-\alpha_n} $,
with $\alpha_n=(9+3n)/(5+n)$. This profile will transit to a 
steeper power law, determined by say secondary infall, 
beyond the core radius. The core radius, which is also the radius,
where the density has an effective logarithmic slope of $-2$, is
related to and a fraction of the turn around radius for the matter
infalling to the halo. The characteristic density at this radius is a
few hundred times the turn around density, and therefore
correlates inversely with typical halo mass, and directly
with the halo formation redshift. 

Although scaling laws suggest a possible form of the core density
profile, they do not tell us how and in fact whether this
form will be realized dynamically. To explore this dynamical issue,
we have adopted two complimentary approaches. First, in section 3,
we studied a simple tractable model: The spherical self similar collapse 
of dark matter density perturbations, in a flat universe.
Then in section 5, we analyzed the properties of halos
obtained in some cosmological N-body simulations,
with a power law spectrum of density fluctuations.

The problem of spherical self similar collapse,
has often been solved earlier by following 
particle trajectories. We adopted instead another approach, 
examining directly the evolution of the moments of
the phase space density. For a purely radial
collapse, with the initial density profile $\propto r^{-3\epsilon}$, 
and steeper than $r^{-2}$, we recover, by demanding that the core be static,
the asymptotic form of the non-linear density profile: 
$\rho \propto r^{-\alpha} \propto r^{-9\epsilon/(1 + 3\epsilon)}$
(see also Padmanabhan 1996b). 
For initial density profiles shallower 
than $1/r^2$, with $\epsilon < 2/3$,
we showed that, non radial velocities are necessarily
required to obtain a static core. These results agree with
the work of B85 and FG  who followed particle trajectories
to solve this problem. 

The consequences of introducing non radial velocity dispersions,
in this approach, can only be examined, by adopting
a closure approximation to the moment equations. 
In the spherical collapse problem, the skewness of
the tangential velocities can be assumed to be zero,
in the core regions. Infact,
in regions where large amounts
of shell crossing has occurred, one can assume that 
a quasi "equilibrium" state obtains,
whereby all odd moments of the distribution function, over 
$({\bf v} - \bar{\bf v})$, may be neglected. 
One can then analytically integrate
the Jeans equation for the self similar collapse, 
including the effect of tangential velocity dispersions.
For an initial density profile shallower that $1/r^2$, 
with $\epsilon < 2/3$, a static core with a non-linear density profile,
with $\alpha= 9\epsilon/(1 + 3\epsilon)$, is
possible, only if the core has sufficiently large tangential
velocity dispersions.  In fact, one
needs $\bar{v_{\theta}^2} > GM/2r$. 
Also if a static core has to have a cuspy density
profile shallower than $1/r$, (with $\alpha < 1$), 
one requires $\bar{v_{\theta}^2} > \bar{v_r^2}$. 
Importantly for the case $3\epsilon = (3+n)/2$ 
(as would be relevant for collapse around a typical point in the 
universe), we recover the simple result $\alpha=\alpha_n
= (9 +3n)/(5 + n) $, with $\alpha_n < 2$, for $n < 1$.

Note that the radial peculiar velocity which could have
negligible skewness in the core, will necessarily have a
non-zero skewness (non zero third moment) near a caustic radius,
where collapsing dark matter particles meet the
outermost shell of re-expanding matter.
In a companion paper (S99),
to take this into account, we have introduced the following 
fluid approach. In this approach, the effect of peculiar 
velocity skewness are neglected in all regions except 
at location of the caustic,
which we call the shock. In the particle picture the shock
is where a single stream flow becomes a muti stream flow.
In the fluid picture it is a where some of the
average infall velocity (of the single stream flow),
is converted to velocity dispersion (of the multi stream flow). 
The location of the caustic, $y_s$, in
scaled co ordinates, is found 
as an eigenvalue to the problem of matching 
the single stream collapse solution at $y > y_s$, with 
a static core solution within $y << y_s$, as determined here. 
This is done in S99 by numerically integrating
the full set of moment equations.
The results largely bear out the expectations of section 3;  
the importance of
tangential velocity dispersions, in deciding the nature of
the core density profile. For the details please see S99. 

The above results and that of S99, although derived for the
case of a purely self similar collapse, illustrate
the importance of dynamical considerations, in determining
the structure of halo cores. They illustrate features which
are like to obtain in more realistic collapse:
If newly collapsing material is constrained to mostly contribute to
the density at larger and larger radii, then memory
of initial conditions can be retained. 
In the more general case, when newly collapsing
material is able to occupy similar regions as the matter
which collapsed earlier, the core density profile 
will only partially reflect a memory of the initial conditions.

The density profiles of the outer regions of dark halos
are briefly studied in Section 4, relaxing the restriction 
to a flat universe and following the Gunn-Gott paradigm. 
For an open universe and at late times, 
the outer density profile goes over to a limiting form
$\rho(r) \propto r^{-4}$, where the slope is independent
of the power law slope of the initial density profile. 
The corresponding
limiting outer profile for a $\Lambda$ dominated universe was shown to
have a form
$\rho(r) \propto [1 - (r/\bar r_{\lambda})^{-3\epsilon}]^{1/2}$, 
where $\bar r_\lambda$ is a characteristic cut-off radius. 
These density profile laws show that in open
and $\Lambda$ dominated models, the halo mass is convergent and
halos have characteristic density cut-off's, which may be 
observationally testable.

We then turned to a complimentary approach, of looking
at halo properties in numerical simulations of
structure formation models 
with a power spectrum $P(k) \propto k^n$; with 3 different values
of the spectral index $n=-2,-1,0$, and with $\Omega_0=1$. 
The results are summarized in figures 2 - 7.
One preliminary conclusion, 
is that the core density profiles of dark matter halos, do depend on 
the initial power spectrum of density fluctuations;
with the local core slope becoming steeper
as the spectral index increases from $n = -2$ to $n =0$.

For example, the median value of the inner core slope $s(r_i)$ 
for the $n=-2$ models is $1.3 \pm 0.07$, 
while for the $n = -1$ and $n= 0$ models, the corresponding
median value of the core slope shifts to $ 1.6 \pm 0.09$, 
and $1.8 \pm 0.09$  respectively. 
For any fixed $n$, there is also a systematic increse of the median slope 
as one increases $r_0$ and goes from the left-most to 
the right most histogram in Figure 5.
These values are generally steeper than $\alpha_n = 1, 1.5$ and $1.8$,
which scaling arguments predict for models with $n=-2,-1$ and $0$
respectively. Further the Kolmogorov-Smirnov two sample test
shows that the distribution
of core slopes of the $n=0$ model is stochastically larger than the
core slopes of the $n=-2$ model, and not drawn from the same population,
at a $ 99\%$ confidence level. It also rules out the hypothesis
that the core slopes of the 
$n=-1$ and $n=-2$ models (or the $n=0$ and $n=-1$ models)
are drawn from the sampe population, at a weaker $90\%$ confidence level. 
The NFW value of $\alpha_0 = 1$ is not favored;
most halos having a steeper core density profile.
Some recent higher resolution simulations, of cluster and 
galactic scale dark halo formation in the CDM model, 
by Moore {\it et al.} (1998, 1999), which resolve the core
very well ( but only for a few halos), has also
obtained a core profile $\rho \propto r^{-1.5}$, 
steeper than the NFW profile.

The velocity dispersion profiles of halo cores, 
in the N-body simulations are somewhat noisy, 
but do indicate for most halos  
a rise in the radial velocity dispersion in the 
core regions. The tangential dispersions are smaller than radial and also
in general show much weaker (sometime no) rise in the core.
The Moore {\it et al.} (1998) cluster halo also shows a rise in the velocity
dispersions in the core (Moore, private communication). 
Indeed Moore {\it et al.} (1998) point out that
a significant fraction of the core material is made from
high density material which collapsed at higher redshift.
It would be very useful to see in the future
whether these trends hold up with large statistical studies  
of halo density profiles, with higher spatial resolution.

An understanding of what determines the core density
profiles of dark halos is important for several issues.
Perhaps one of the most
relevant and dramatic effect will be on strong gravitational
lensing properties of galactic and clusters scale dark matter halos.
For example the multiple imaging lensing cross
sections will be very different whether one has
$\alpha=0, -1$ or $\alpha=-2$. Compare the work of
Subramanian, Rees and Chitre (1987) where 
the lensing cross section by dark halos 
was estimated assuming that they have soft cores with the work of
Narayan and White (1988) where halos were assumed
to be singular isothermal spheres.

More specifically, a given system is capable of producing multiple
images if its surface density exceeds a critical value
$\Sigma_c$ (Turner, Ostriker and Gott 1984; Subramanian and Cowling 1986).
For $\alpha \geq 1$ the surface density is divergent in the 
central regions. So, generically, if $\alpha \geq 1$,
{\it all} clusters are capable of producing multiple images.
For $\alpha < 1$, some clusters can produce multiple images
and some can not. The NFW value of $\alpha = 1$
is at the boundary with surface density logarithmically singular.
Thus the actual values of $\alpha$ for real systems 
is quite relevant to the frequency of strong lensing. 
As the frequency of multiple images from gravitational lensing also
provides a powerful independent test of cosmological models,
(cf. Cen {\it et al.}, 1994, Wambsganss, Cen and Ostriker 1998,
Cen 1998), it is important to determine the structure of halo cores.
Further, the existence and properties of radial arcs, which
have been observed in some cluster
scale lenses depends on the slope of the inner cusp (cf. 
Mellier, Fort and Kneib 1993, 
Miralda-Escude 1995, Bartelmann 1997, Evans and Wilkinson 1998). 
For the singular isothermal profile and stronger cusps,
radial arcs do not form. 

The rotation curves of disk galaxies 
also holds clues to the core density profile of dark halos. 
But it is more difficult to decompose the observed rotation curve
unambiguously into contributions from the luminous stellar disk/bulge
and the dark halo.
The fluid approach adopted in section 3 and in S99 raises a
new way of exploring non linear dynamics, which
can extend analytic approximations like the Zeldovich
approximation, valid in a single stream flow, to
the multi streaming regime. In the fluid approach multistreaming
regions would correspond to regions with velocity dispersions,
generated by the Zeldovich type caustics. Note that the adhesion
approximation, is one extreme where the multi streaming
regions are collapsed onto a caustic. It would be interesting
to explore this issue further.
In this work we have not included
the dynamics of the gaseous (the baryonic)
component, which will be in fact relevant for the interpretation of
x-ray observations of clusters. The gas necessarily has an isotropic
velocity dispersion, and so will have a different dynamical
evolution compared to the dark matter. We hope to return to
some of these issues in the future.

\acknowledgments

This work was begun when KS visited the Princeton University
Observatory, during Sept-Nov 1996. 
Partial travel support to Princeton came from IAU Commission 38.
Some of the work was done at the University of Sussex where 
KS was supported by a PPARC
Visiting Fellowship. He thanks John Barrow, Ed Turner, the other 
Princeton and Sussex astronomers for warm hospitality.
T. Padmanabhan is thanked for critical comments on
an earlier version of this work. KS also thanks
Ben Moore, Bepi Tormen, Ravi Sheth, Dave Syer
and Simon White for several helpful discussions. 
This research is supported in part by grants AST93-18185
and ASC97-40300.

\clearpage
 
%
%

\clearpage

\begin{figure*}
\centering
\begin{picture}(400,250)
\psfig{figure=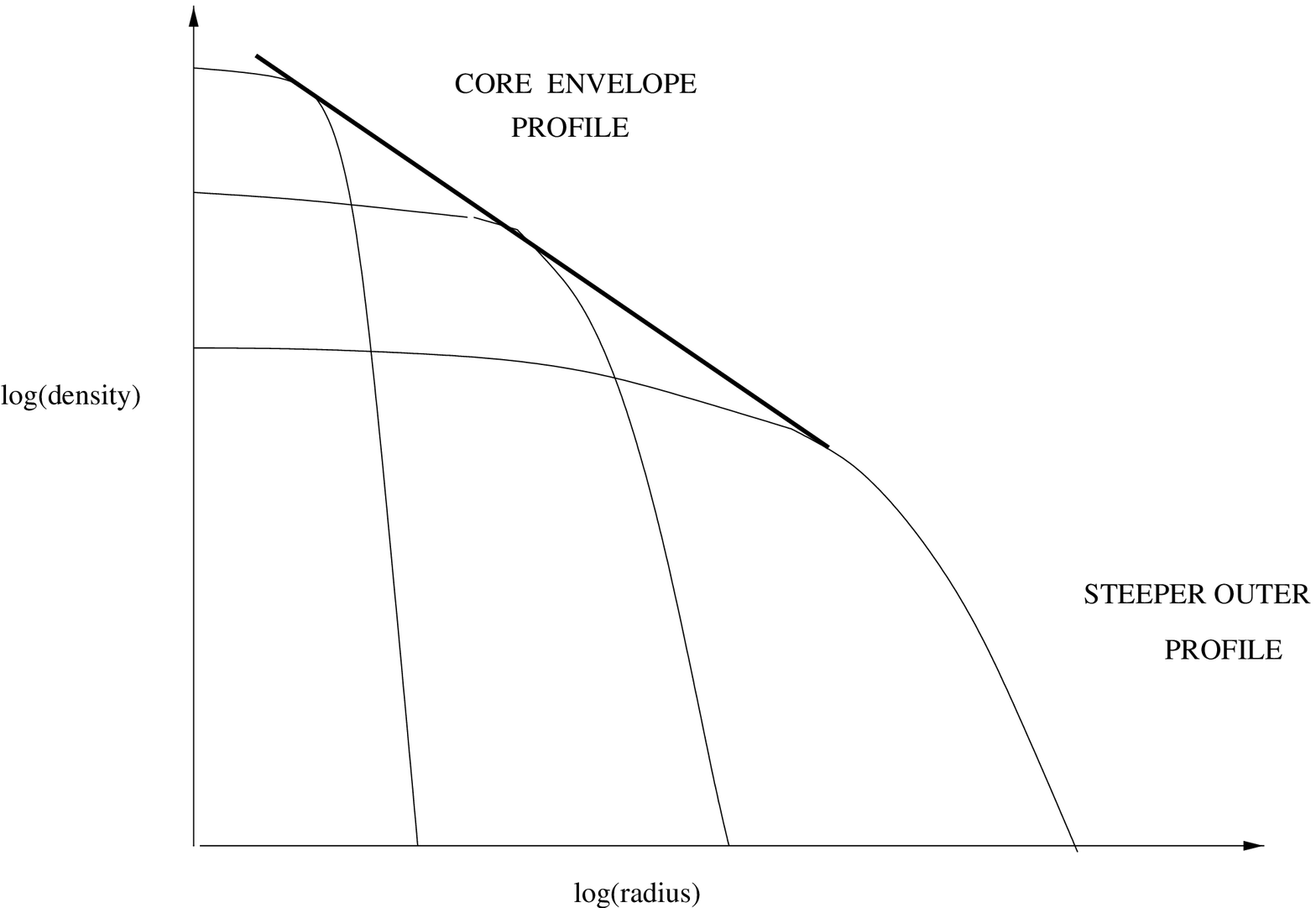,height=13.0cm,width=15.0cm,angle=0.0}
\end{picture}
\caption{ Schematic illustration of how the density
profile of a large halo core could arise as an envelope of the density
profiles of undigested cores of smaller mass halos.}
\end{figure*}

\clearpage

\begin{figure*}
\centering
\begin{picture}(600,550)
\psfig{figure=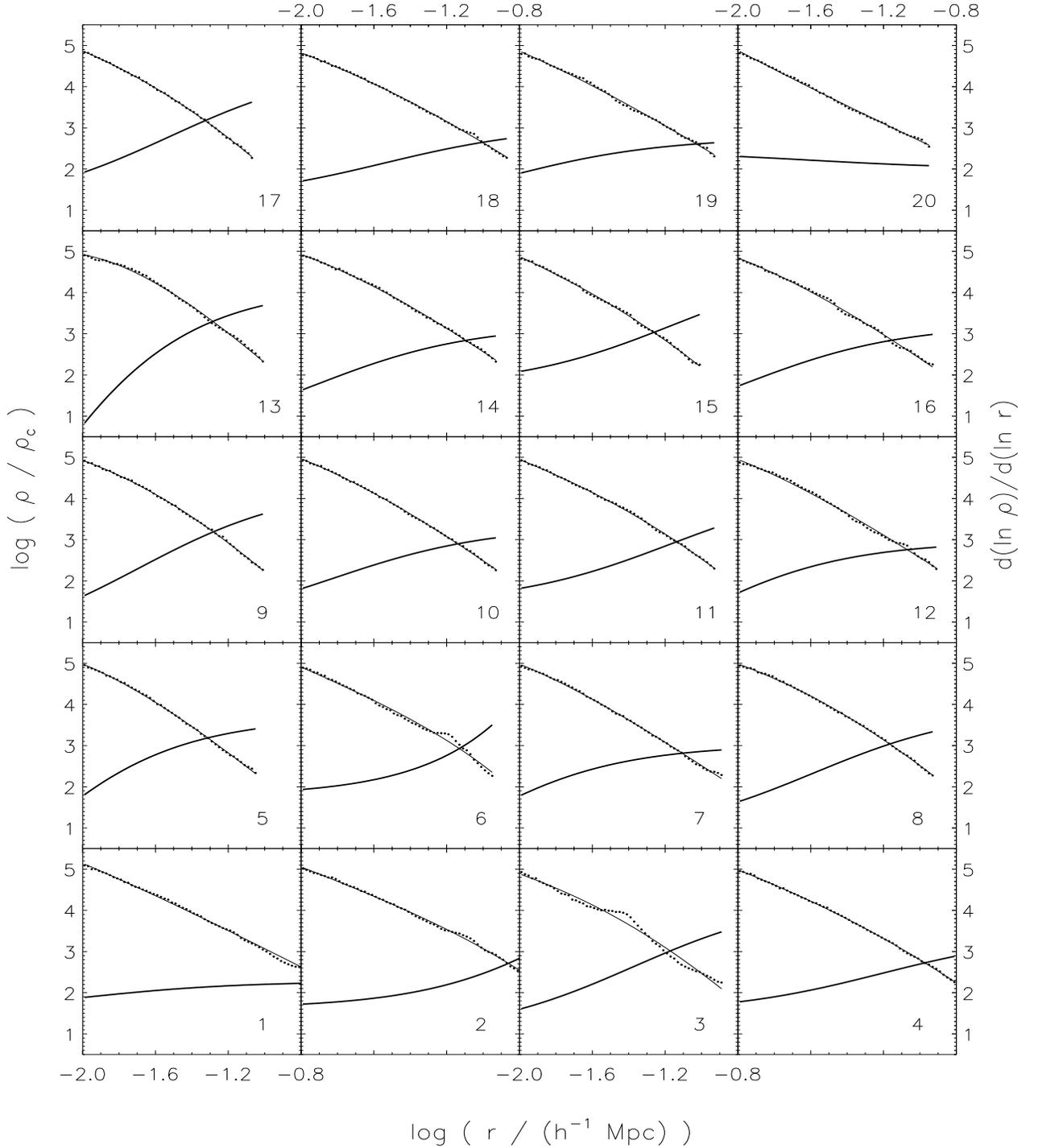,height=20.0cm,width=18.0cm,angle=0.0}
\end{picture}
\caption{ The density profiles of the 20 halos 
in N-body simulations with $n=0$, are shown as dotted
lines in each panel. The halos are numbered in
each panel for easy identification. 
A double power law fit to each density profile data 
is superposed as a light solid line in these figures.
For each halo, the local logarithmic slope of the density profile,
$s(r) = d(ln \rho)/d(ln r)$, calculated from the model fit
is shown as a thick solid line, in the same plot as the density profile
plot. 
 }
\end{figure*}

\clearpage

\begin{figure*}
\centering
\begin{picture}(600,550)
\psfig{figure=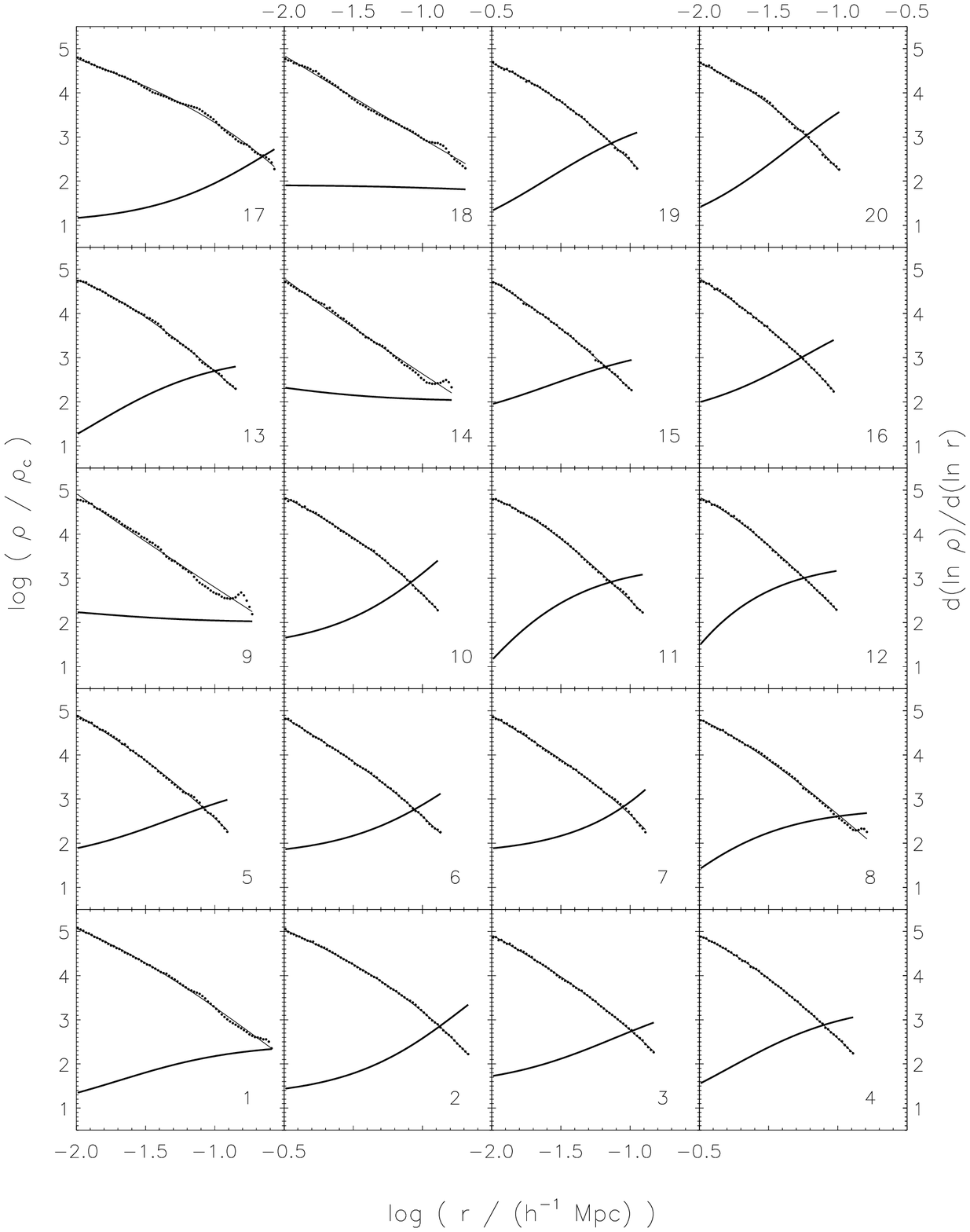,height=20.0cm,width=18.0cm,angle=0.0}
\end{picture}
\caption{The density profiles and the logarithmic slopes
for the 20 halos in the simulation with $n = -1$.
The dotted line, thin solid line and thick solid line as as in Fig. 2.
 }
\end{figure*}

\clearpage

\begin{figure*}
\centering
\begin{picture}(600,550)
\psfig{figure=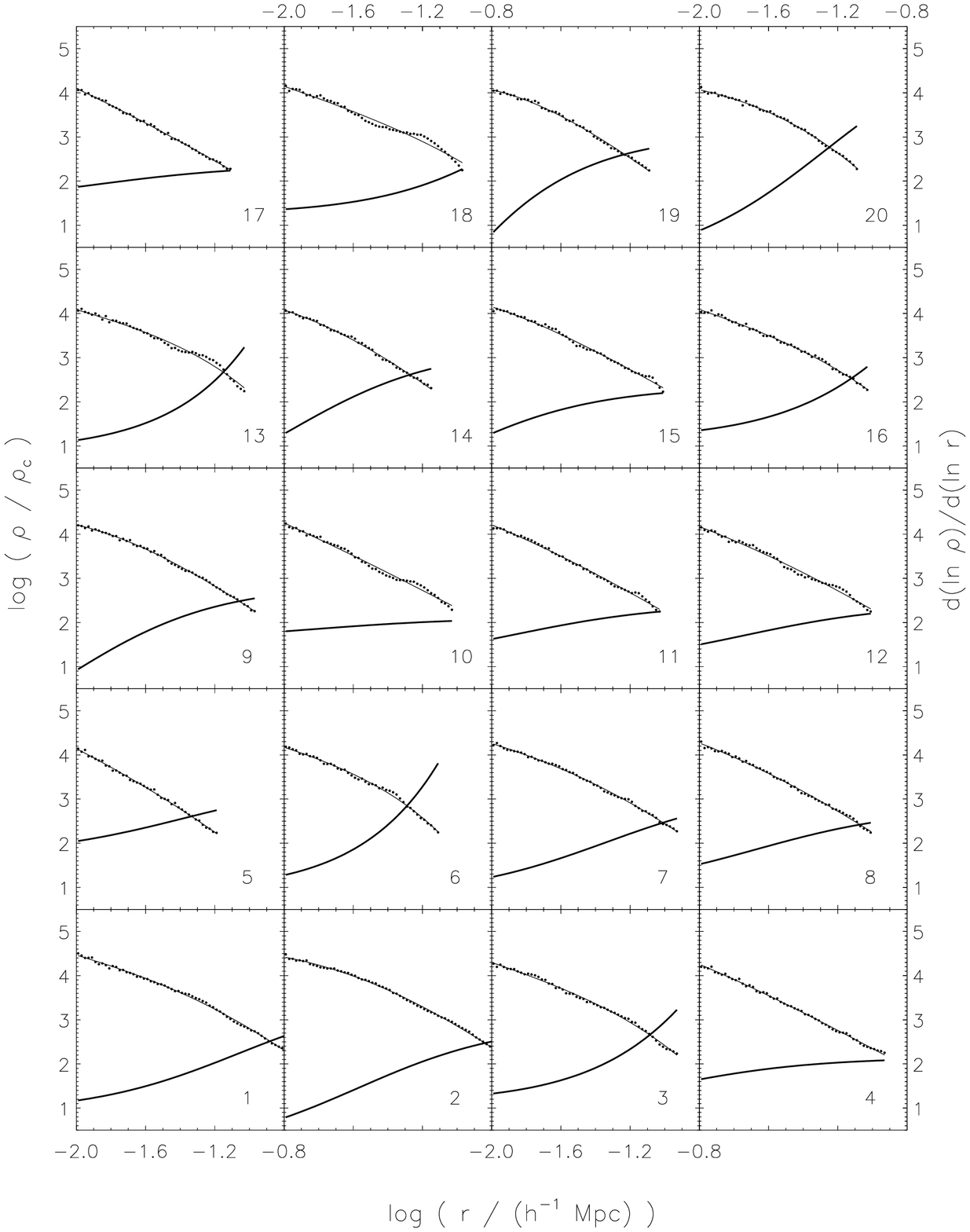,height=20.0cm,width=18.0cm,angle=0.0}
\end{picture}
\caption{The density profiles and the logarithmic slopes
for the 20 halos in the simulation with $n = -2$.
The dotted line, thin solid line and thick solid line as as in Fig. 2.
 }
\end{figure*}

\clearpage

\begin{figure*}
\centering
\begin{picture}(650,550)
\psfig{figure=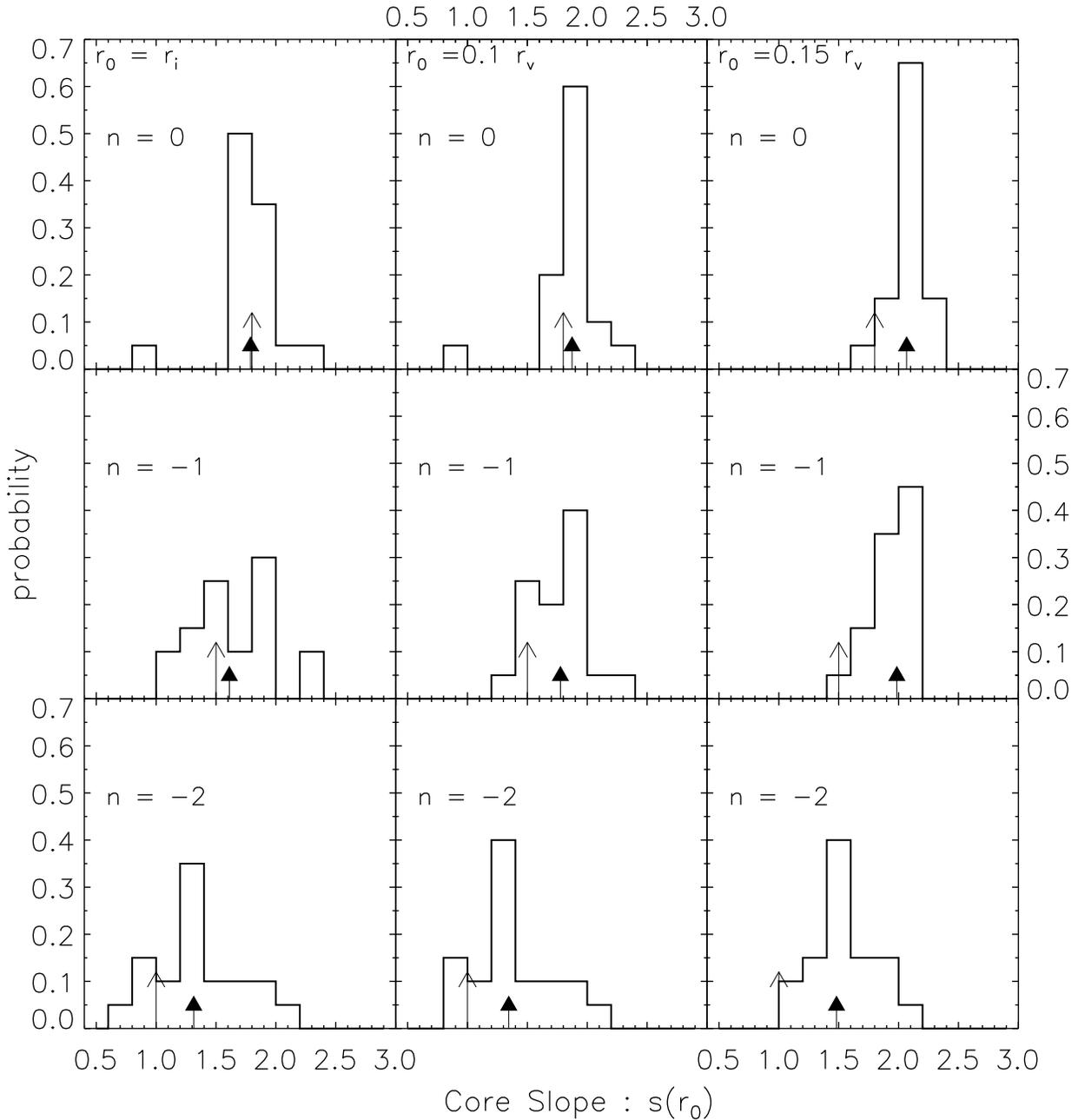,height=18.0cm,width=17.0cm,angle=0.0}
\end{picture}
\caption{ The distribution of core slopes of halo density profiles
in different models. The solid arrow shows the location of the
median of the distribution while the thin arrow indicates
the location of the value $3(3+n)/(5+n)$, predicted by scaling
arguments.
 }
\end{figure*}

\clearpage

\begin{figure*}
\centering
\begin{picture}(600,550)
\psfig{figure=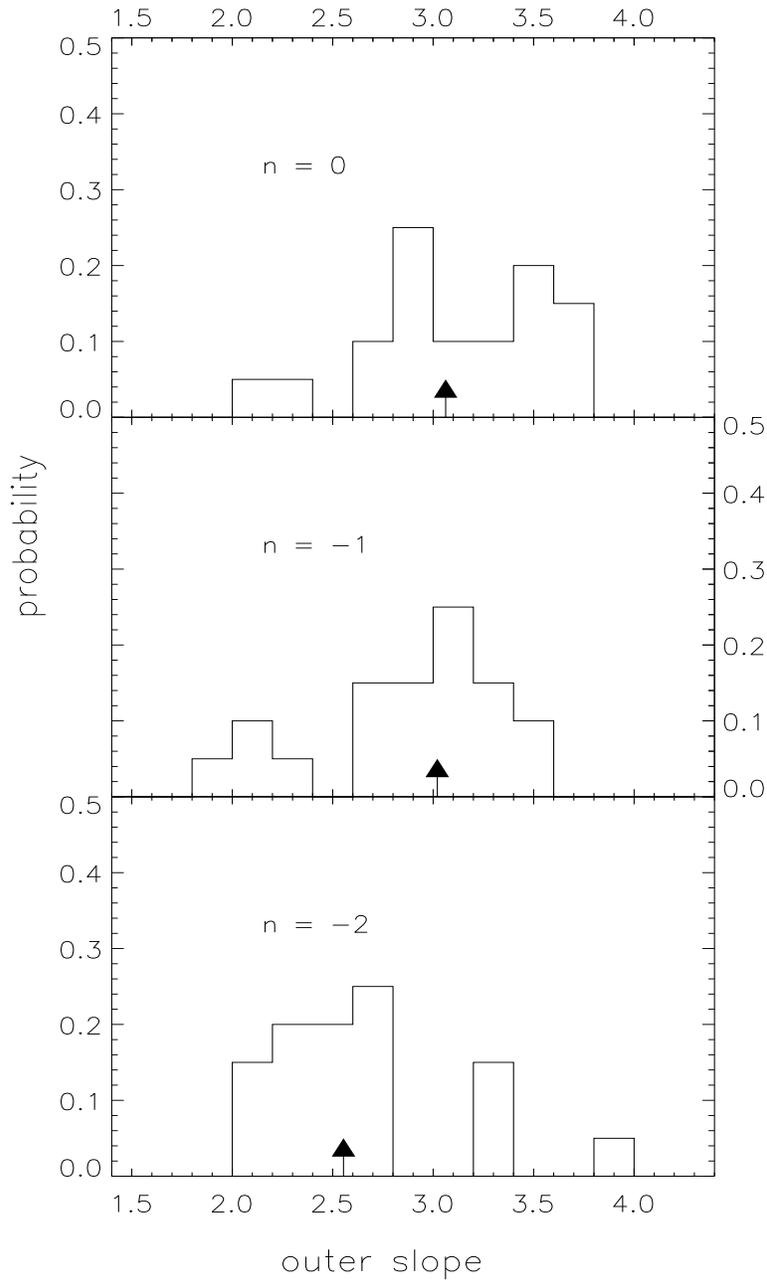,height=18.0cm,width=20.0cm,angle=0.0}
\end{picture}
\caption{ The distribution of outer slopes of halo density profiles
in different models. The solid arrow shows the location of
the median of the distribution.
 }
\end{figure*}

\clearpage

\begin{figure*}
\centering
\begin{picture}(600,550)
\psfig{figure=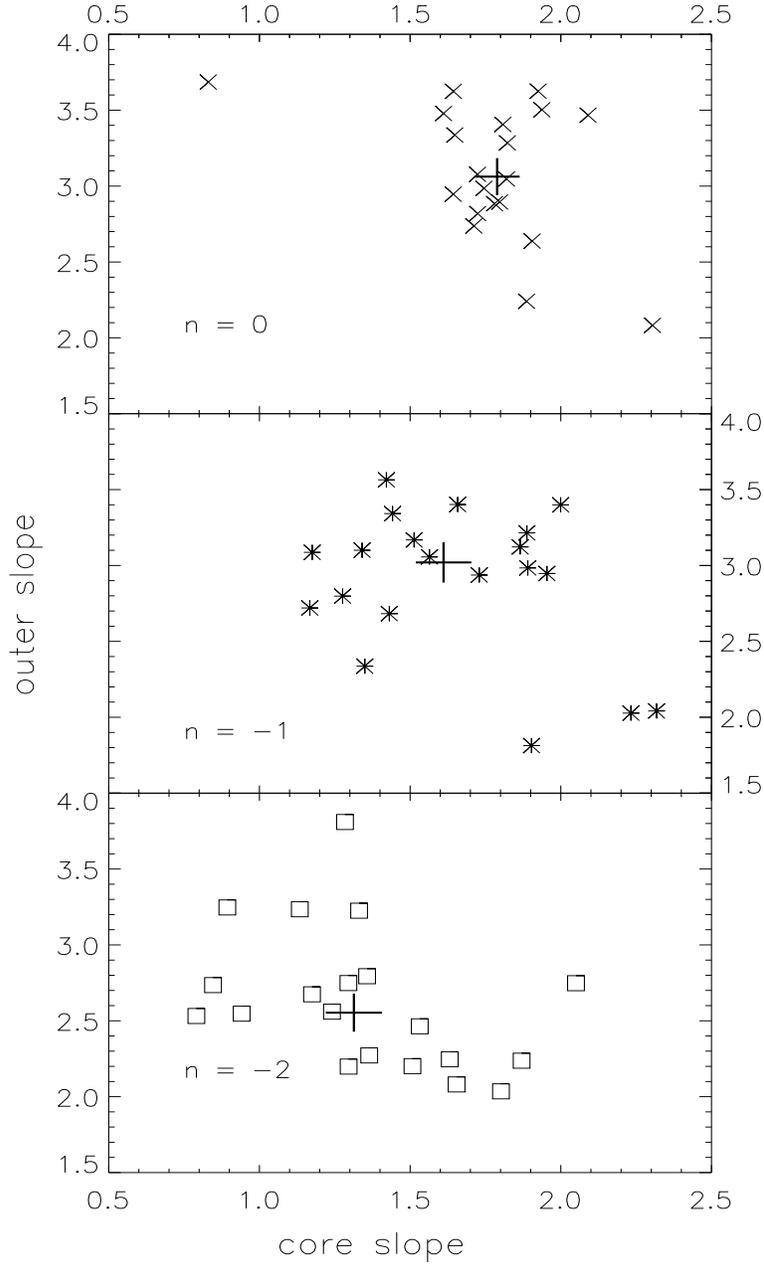,height=18.0cm,width=20.0cm,angle=0.0}
\end{picture}
\caption{ The distribution of the inner core and outer slopes
 of halo density profiles in different models. The cross marks
the median value of these slopes and its extent  
gives the $\pm$ error on the median values. 
 }
\end{figure*}

\end{document}